\documentclass[twocolumn,showpacs,aps,prl,superscriptaddress]{revtex4}
\usepackage{dcolumn}
\usepackage{amsmath}
\usepackage{epsfig}
\usepackage{feynmp}
\usepackage{graphicx}
\usepackage{lineno}

\newcommand{\BaBarYear}       {05}
\newcommand{\BaBarNumber}     {14}
\newcommand{\SLACPubNumber} {11255}

\newcommand{\BaBarType}      {PUB}  

\RequirePackage{xspace}

\usepackage{relsize}
\def\babar{\mbox{\slshape B\kern-0.1em{\smaller A}\kern-0.1em
    B\kern-0.1em{\smaller A\kern-0.2em R}}}

\def\qqbar {\ensuremath{q\overline q}\xspace}

\def\piz   {\ensuremath{\pi^0}\xspace}

\def\pip   {\ensuremath{\pi^+}\xspace}
\def\pim   {\ensuremath{\pi^-}\xspace}

\def\Kbar  {\kern 0.2em\overline{\kern -0.2em K}{}\xspace}

\def\Kz    {\ensuremath{K^0}\xspace}
\def\Kzb   {\ensuremath{\Kbar^0}\xspace}
\def\KzKzb {\ensuremath{\Kz \kern -0.16em \Kzb}\xspace}
\def\Kp    {\ensuremath{K^+}\xspace}
\def\Km    {\ensuremath{K^-}\xspace}

\def\KpKm  {\ensuremath{\Kp \kern -0.16em \Km}\xspace}
\def\KS    {\ensuremath{K^0_{\scriptscriptstyle S}}\xspace}

\def\Dbar    {\kern 0.2em\overline{\kern -0.2em D}{}\xspace}

\def\Dz      {\ensuremath{D^0}\xspace}
\def\Dzb     {\ensuremath{\Dbar^0}\xspace}
\def\DzDzb   {\ensuremath{\Dz {\kern -0.16em \Dzb}}\xspace}
\def\Dp      {\ensuremath{D^+}\xspace}
\def\Dm      {\ensuremath{D^-}\xspace}

\def\DpDm    {\ensuremath{\Dp {\kern -0.16em \Dm}}\xspace}

\def\Dstarp  {\ensuremath{D^{*+}}\xspace}

\def\Bbar    {\kern 0.18em\overline{\kern -0.18em B}{}\xspace}

\def\BB      {\ensuremath{B\Bbar}\xspace} 
\def\Bz      {\ensuremath{B^0}\xspace}
\def\Bzb     {\ensuremath{\Bbar^0}\xspace}
\def\BzBzb   {\ensuremath{\Bz {\kern -0.16em \Bzb}}\xspace}
\def\Bu      {\ensuremath{B^+}\xspace}
\def\Bub     {\ensuremath{B^-}\xspace}
\def\Bp      {\ensuremath{\Bu}\xspace}

\def\BpBm    {\ensuremath{\Bu {\kern -0.16em \Bub}}\xspace}

\def\BorBbar    {\kern 0.18em\optbar{\kern -0.18em B}{}\xspace}
\def\DorDbar    {\kern 0.18em\optbar{\kern -0.18em D}{}\xspace}
\def\KorKbar    {\kern 0.18em\optbar{\kern -0.18em K}{}\xspace}

\mathchardef\Upsilon="7107
\def\Y#1S{\ensuremath{\Upsilon{(#1S)}}\xspace}

\def\FourS {\Y4S}

\mathchardef\Deltares="7101
\mathchardef\Xi="7104
\mathchardef\Lambda="7103
\mathchardef\Sigma="7106
\mathchardef\Omega="710A

\def\Deltabar{\kern 0.25em\overline{\kern -0.25em \Deltares}{}\xspace}
\def\Lbar{\kern 0.2em\overline{\kern -0.2em\Lambda\kern 0.05em}\kern-0.05em{}\xspace}
\def\Sigbar{\kern 0.2em\overline{\kern -0.2em \Sigma}{}\xspace}
\def\Xibar{\kern 0.2em\overline{\kern -0.2em \Xi}{}\xspace}
\def\Obar{\kern 0.2em\overline{\kern -0.2em \Omega}{}\xspace}
\def\Nbar{\kern 0.2em\overline{\kern -0.2em N}{}\xspace}
\def\Xb{\kern 0.2em\overline{\kern -0.2em X}{}\xspace}

\def\X {\ensuremath{X}\xspace}

\def\mes        {\mbox{$m_{\rm ES}$}\xspace}

\newcommand{\tev}{\ensuremath{\mathrm{\,Te\kern -0.1em V}}\xspace}
\newcommand{\gev}{\ensuremath{\mathrm{\,Ge\kern -0.1em V}}\xspace}
\newcommand{\mev}{\ensuremath{\mathrm{\,Me\kern -0.1em V}}\xspace}
\newcommand{\kev}{\ensuremath{\mathrm{\,ke\kern -0.1em V}}\xspace}
\newcommand{\ev}{\ensuremath{\mathrm{\,e\kern -0.1em V}}\xspace}
\newcommand{\gevc}{\ensuremath{{\mathrm{\,Ge\kern -0.1em V\!/}c}}\xspace}
\newcommand{\mevc}{\ensuremath{{\mathrm{\,Me\kern -0.1em V\!/}c}}\xspace}
\newcommand{\gevcc}{\ensuremath{{\mathrm{\,Ge\kern -0.1em V\!/}c^2}}\xspace}
\newcommand{\mevcc}{\ensuremath{{\mathrm{\,Me\kern -0.1em V\!/}c^2}}\xspace}

\def\invfb   {\ensuremath{\mbox{\,fb}^{-1}}\xspace}

\def\mus  {\ensuremath{\rm \,\mus}\xspace}

\def\mus        {\ensuremath{\,\mu{\rm s}}\xspace}    

\def\to                 {\ensuremath{\rightarrow}\xspace}

\def\pep2{PEP-II}

\def\gsim{{~\raise.15em\hbox{$>$}\kern-.85em
          \lower.35em\hbox{$\sim$}~}\xspace}
\def\lsim{{~\raise.15em\hbox{$<$}\kern-.85em
          \lower.35em\hbox{$\sim$}~}\xspace}

\xspace

\newcommand{\jprlBase}       {Phys.\ Rev.\ Lett.\xspace}

\newcommand{\zpBase}         {Z.\ Phys.\xspace}

\newcommand{\jprl}      [1]  {\jprlBase\ {\bf #1}}

\newcommand{\zpc}       [1]  {\zpBase\ C~{\bf #1}}

\def\jetset74   {\mbox{\tt Jetset \hspace{-0.5em}7.\hspace{-0.2em}4}\xspace}

\newcommand{\bchdstarcks}{$B^+ \rightarrow  D^{(*)+} K^0_{s}$}
\newcommand{\nbb}{226}
\newcommand{\dclim}{ 0.5 \times 10^{-5}}
\newcommand{\dstarclim}{0.9 \times 10^{-5}}
\def\myprl  #1 #2 #3 {\jprl{#1},\ #2 (#3)}

\begin{document}

\begin{flushleft}
\babar-\BaBarType-\BaBarYear/\BaBarNumber \\
SLAC-PUB-\SLACPubNumber
\end{flushleft}
\noindent
\title{
{\large \bf Search for the Rare Decays  \boldmath{ \bchdstarcks} } 
}

%
\author{B.~Aubert}
\author{R.~Barate}
\author{D.~Boutigny}
\author{F.~Couderc}
\author{Y.~Karyotakis}
\author{J.~P.~Lees}
\author{V.~Poireau}
\author{V.~Tisserand}
\author{A.~Zghiche}
\affiliation{Laboratoire de Physique des Particules, F-74941 Annecy-le-Vieux, France }
\author{E.~Grauges}
\affiliation{IFAE, Universitat Autonoma de Barcelona, E-08193 Bellaterra, Barcelona, Spain }
% ...
\collaboration{The \babar\ Collaboration}
\noaffiliation

\begin{abstract}
We report on the search for the rare decays \bchdstarcks\  in 
 approximately \nbb\ ~million~ \Y4S~$\to$~ \BB decays collected with the
\babar\ detector at the PEP-II asymmetric-energy $B$ factory at SLAC.
We do not observe any   significant signal and we set 90\% confidence level upper limits on the branching fractions, 
${\ensuremath {{\cal B}(\Bp{\to}D^{+}K^0})}$ $<\dclim$ and
${\ensuremath {{\cal B}(\Bp{\to}D^{*+}K^0})}$ $<\dstarclim$.
\end{abstract}


\maketitle

\begin{figure*}[bht]
\begin{center}
\epsfig{file=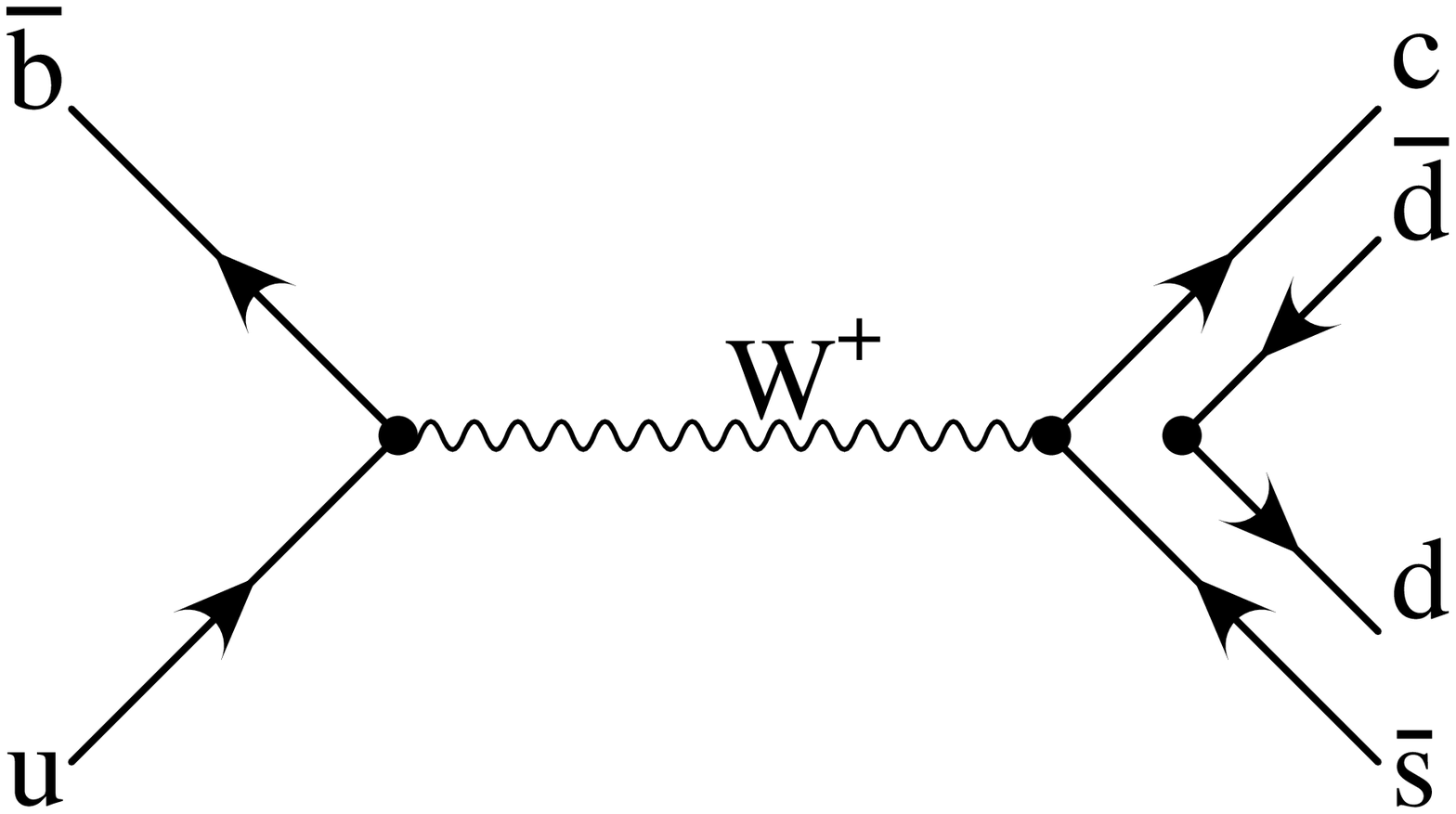,height=2.5cm}
\epsfig{file=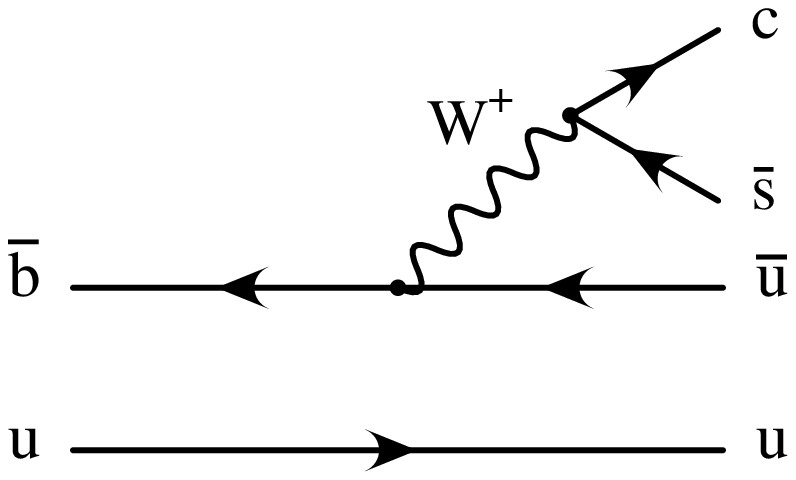,height=3cm}
\epsfig{file=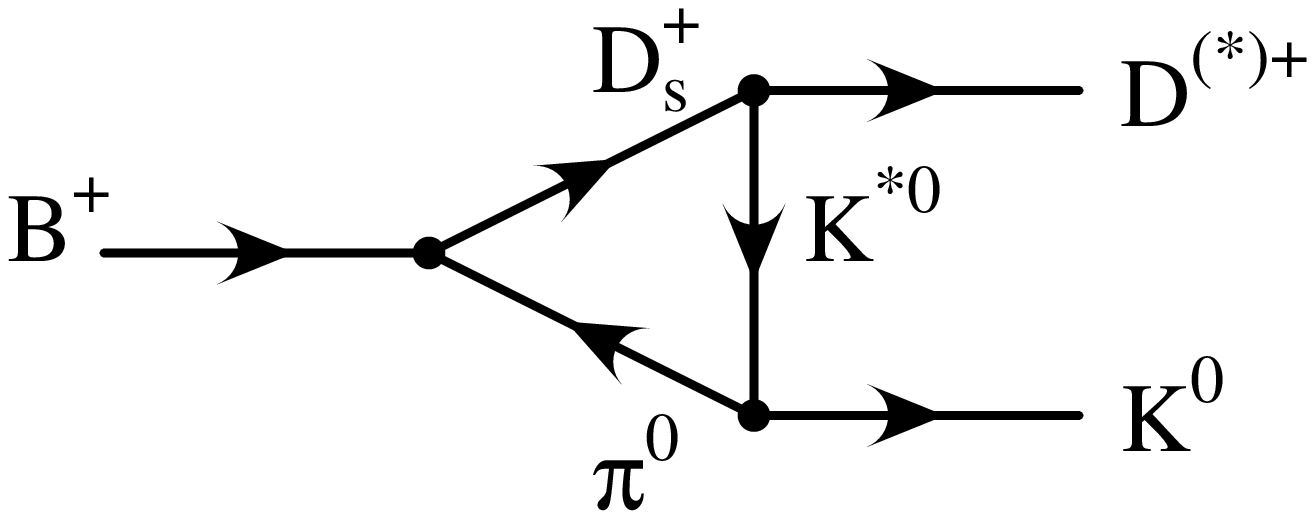,height=3cm}
\end{center}
\caption{ Annihilation diagram for the decay ${\ensuremath {\Bp{\to}D^{(*)+}\KS}}$(left), tree diagram for ${\ensuremath {\Bp{\to}D_s^{(*)+}\piz}}$(center), and  hadron-level diagram for a possible rescattering contribution to ${\ensuremath {\Bp{\to}D^{(*)+}\KS}}$ via  ${\ensuremath {\Bp{\to}D_s^{(*)+}\piz}}$(right).}
\label{fig:feyn2}
\end{figure*}

Meson decays in which neither constituent quark appears in the final
state are expected to be dominated by annihilation diagrams, in which
the two quarks interact directly. Such processes
provide interesting insights into the internal dynamics of $B$ mesons 
and need to be understood to make precise 
predictions on $B$ meson decays.
Such diagrams cannot be calculated by assuming factorization since both the quarks play a
 role and a reliable theoretical prediction for the corresponding amplitudes does not exist.
These amplitudes are expected to be suppressed with respect to  amplitudes 
where one of the two quarks is a spectator by
 a factor $\sim {{\rm f}_B}/{{\rm m}_B}\sim 0.04$ (${\rm f}_B\sim 200 \mev$ 
and ${\rm m_B}= 5.28 \gevcc$ are the $B$ meson decay 
constant~\cite{CERNyellow} and mass, respectively). This factor represents the  amplitude for the two
 quark wave functions overlapping, a necessary condition in annihilations.
 So far no process relying entirely on annihilation has  been observed and the assumption that these types of diagrams can be 
neglected is frequently used in theoretical calculations.
Some studies~\cite{Blok:1997yj} indicate, though, that processes with a spectator quark can contribute to annihilation-mediated decays 
by {\it {rescattering}} if the final state is reached in two
steps: a decay into two mesons that can occur with a spectator quark, and a subsequent strong interaction
 between the two mesons which produces the final state of interest. Figure~\ref{fig:feyn2} shows the Feynman
 diagram for the decays ${\ensuremath {\Bp{\to}D^{(*)+}K^0_s}}$ and ${\ensuremath {\Bp{\to}D_s^{+}\pi^0}}$~\cite{ch},
 and the hadron-level diagram for the  rescattering. Strong rescattering could then mimic large contributions from annihilation diagrams
to the level of not being negligible any more.

The decays ${\ensuremath {\Bp{\to}D^{(*)+}K^0_s}}$ are particularly suited to study annihilations because
 of their relatively clean experimental signature
 and because their branching fractions are expected to be at the level of the current sensitivity ($10^{-5}$) 
if large rescattering occurs, or three orders of magnitude below if not~\cite{Blok:1997yj}.
Moreover the branching fraction of these decays can be used to constrain the annihilation amplitudes in 
 the phenomenological fits~\cite{buras} that  allows to translate 
the measurement of the amplitude of ${\ensuremath {B^+{\to}D^{0}K^+}}$  into 
estimates of the Cabibbo-suppressed decay
${\ensuremath {B^0{\to}D^{0}K^0}}$ needed in some CP measurements~\cite{d0k0}.
Neither of the modes studied here has been
observed so far, and a 90$\%$ confidence level  upper limit on the  branching fraction  
${\ensuremath {{\cal B}(\Bp{\to}D^{*+}K^0})} < 9.5\times10^{-5}$  has been established by CLEO~\cite{prior}.

In this paper we present the results of the search for \bchdstarcks\ decays in $225.9\pm2.5$ million \Y4S $\to$ \BB decays, collected 
with the \babar\ detector~\cite{detector} at the PEP-II asymmetric-energy $B$ factory at SLAC. 
We use a Monte Carlo (MC) simulation of the \babar\ detector based on
GEANT4~\cite{geant} to validate the analysis procedure, estimate efficiencies, and to study the relevant backgrounds. We also
use 12.4 \invfb of data collected at a center-of-mass energy approximately 40\mev below the \FourS mass.
 
Candidates for \Dp mesons are reconstructed in the modes $D^+ \rightarrow K^-\pi^+\pi^+$ and
$D^+ \rightarrow K^0_s\pi^+$. Candidates for \Dstarp mesons are reconstructed in
the mode $\Dstarp \rightarrow \Dz\pip$, where the \Dz subsequently decays to one of the four modes $K^{-}\pi^{+}$, $K^{-}\pi^{+}\piz$, $K^{-}\pi^{+}\pi^{-}\pi^{+}$, or $\KS\pi^{+}\pi^{-}$.

$K^0_s$ candidates are reconstructed from 
two oppositely-charged tracks with an invariant mass $491 < m_{\pip \pim} < 504 \mevcc$ (corresponding to a $\pm2$ standard deviations, $\sigma$, window around the mean value in control samples).
 The $\chi^2$ of the $\pi^+\pi^-$ vertex fit must have a probability greater than 0.1$\%$ and the $K^0_s$ flight distance from the primary vertex in the plane transverse to the beam axis in the event must be greater than 2 mm. Kaons and pions coming from the $D$ are required to have momentum in the laboratory frame greater than 200 $\mevc$ and 150 $\mevc$, respectively, except in the decays 
$\Dz \rightarrow K^-\pi^+$ ($K^-\pi^+\pi^0$) where the momentum threshold for both the tracks is 200 (150) $\mevc$.
To identify charged kaons we use a selection with an efficiency of 95$\%$ and a 
12$\%$ pion misidentification probability. $\pi^0$ candidates are reconstructed combining two photons with invariant mass
$120 < m_{\gamma \gamma} <150 \mevcc$ (corresponding to a $\pm2\sigma$ window around the mean value estimated on
control samples) and a minimum  total energy in the laboratory frame of  200 MeV.
For the $\Dz \rightarrow K^-\pi^+\pi^0$ decay we  select the dominant resonant contributions
 with a requirement on the Dalitz density distribution~\cite{dalitz}.

Finally, the \Dp and \Dz candidates are required to have 
an invariant mass within 2$\sigma$ of the mean values. The \Dp and \Dz mass resolutions are mode-dependent and range between 5 and 8 $\mevcc$.  We form \Dstarp candidates by combining \Dz candidates with charged tracks. The mass difference between the \Dstarp and the \Dz candidates 
 is required to be within 2$\sigma$  of the mean value as estimated on control samples. The resolution  is mode-dependent, approximately 0.6$\mevcc$ in all cases.
We combine \Dp\ or \Dstarp candidates with a $K^0_s$ to form \Bp\ candidates. To improve the resolution on the four-momentum of all the
 intermediate  composite particles we apply a kinematic fitting technique that constrains their masses to the nominal value~\cite{pdg}
and  their charged daughters to come from the same vertex.

\begin{figure}[!htb]
\begin{center}
\includegraphics[width=8cm]{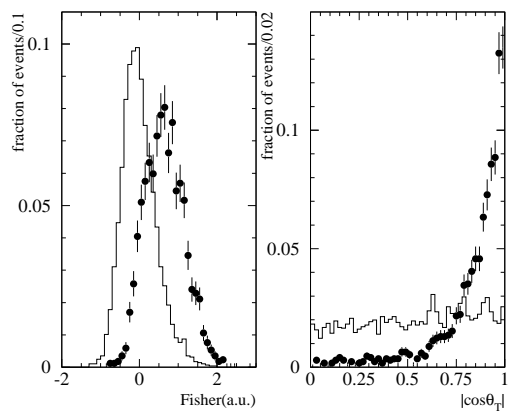}
\caption{Distribution of the discriminating variables   $|$cos$\theta_T$$|$ and ${\cal{F}}$ 
in the  ${\ensuremath {\Bp{\to}D^{+}K^0_s}}$ signal MC (histograms) and the off-resonance data (dots).
\label{fig:vars}
}
\end{center}
\end{figure}

We only accept events with a reconstructed candidate and 
a total measured energy greater than 4.5 GeV, determined
using all charged tracks and neutral clusters in the electromagnetic calorimeter with energy above 30 MeV.
The remaining background comes predominantly from continuum \qqbar\ production.
This background is suppressed using variables that characterize the topology of the event.
We require the ratio of the second
and zeroth order Fox-Wolfram moments~\cite{fox} to be less than 0.5.
We compute the angle $\theta_T$ between the thrust axis of 
 the  $B$-meson candidate and the thrust axis of the rest of the event. The thrust axis is defined as the direction that maximizes the sum of the longitudinal momenta of the particles 
in the center-of-mass (c.m.) frame.
In this frame \BB\ pairs are
produced approximately at rest and have a uniform $|$cos$\theta_T$$|$ distribution.
In contrast, \qqbar\ pairs are produced back-to-back,
which results in a $|$cos$\theta_T$$|$ distribution that peaks at unity.
 To further suppress backgrounds we use a Fisher discriminant $\cal{F}$ constructed from
the scalar sum of the c.m. momenta of all tracks and photons, excluding the $B$ candidate
decay products,  flowing into nine concentric cones centered on the thrust axis of the $B$ candidate~\cite{twobody}. The more spherical the
event, the lower the value of ${\cal{F}}$.
Figure~\ref{fig:vars} shows the distribution of   ${\cal{F}}$ and $|$cos$\theta_T$$|$  on signal MC and 
on off-resonance data, which contain exclusively continuum \qqbar\ production events.
We also exploit the charge correlation between the $B$ and the leptons and kaons produced in its decays
to classify the events in three  mutually exclusive categories with different levels of contamination
from continuum background: events with at least one lepton with  charge opposite to the $B$ candidate, 
events with no lepton and at least one kaon among the tracks that do not
form the $B$ candidate but have opposite charge,
 and all the other events. 
The optimization of the selection is performed separately for each decay mode and for 
the three categories by maximizing the ratio of signal 
efficiency, estimated with MC, over the square-root of
 the expected number of background events, estimated in data sidebands:
the maximum allowed value of  $|$cos$\theta_T$$|$ 
 ranges between 0.8 to 1 (i.e. no cut) and the maximum allowed value  for 
${\cal{F}}$  varies from 0.1 to 0.7.

We extract the signal using the kinematic variables
$m_{ES}= \sqrt{E_{\rm b}^{*2} - (\sum_i {\bf p}^*_i)^2}$
and $\Delta E= \sum_i\sqrt{m_i^2 + {\bf p}_i^{*2}}- E_{\rm b}^*$,
where $E_{\rm b}^*$ is the beam energy in the c.m. frame,
${\bf p}_i^*$ is the c.m.~momentum of daughter particle $i$ of the
$B$ meson candidate, and $m_i$ is the mass hypothesis for particle $i$. 
For signal events, $m_{\rm ES}$ peaks at the $B$ meson mass with 
a resolution of about 2.5 MeV$/c^2$ and $\Delta E$ peaks near zero,  
indicating that the $B$ candidate's total energy is consistent with the beam energy in the c.m.~frame.
The $\Delta E$ signal band is defined as $|\Delta E|<2.5\sigma$ and within it
we define the signal region as $5.2725<m_{ES}<5.2875$ $\gevcc$ 
and the $m_{\rm ES}$ sideband region as $5.2000<m_{ES}<5.2725$ $\gevcc$. The
$\Delta E$ resolution $\sigma$ is mode-dependent and approximately 18 MeV.
We also define the $\Delta E$ sideband region as $2.5\sigma<|\Delta E|<0.12$ $\gev$ and $5.2<m_{ES}<5.3$ $\gevcc$.
Table~\ref{tab:eff} shows the efficiency for each sub-decay mode estimated with simulated events. Depending on the mode, in 1.5 to 7\% of the 
events there is more than one $B$ candidate. We select the B candidate whose $D^{(*)}$ candidate's mass is closest to its
nominal  mass or, in case two B candidates are formed by the same the  $D^{(*)}$ candidate, 
one with the smallest value of $|\Delta E|$.
\begin{table}[t!]
\caption{Efficiencies for the  \bchdstarcks candidate reconstruction in each sub-decay mode. 
 The branching fraction of the $D^{(*)+}$  decay chains considered~\cite{pdg} are also shown.}
\begin{center}
\mbox{ \small
\begin{tabular}{l|c|c} \hline \hline
$D$ mode& $\varepsilon_i$($\%$) &{\cal B}($\%$) \\
\hline
$D^+ \rightarrow \KS\pi^+$; $\KS\to\pim\pip$& 17.3 & 0.97 $\pm$ 0.06\\
$D^+ \rightarrow K^-\pi^+\pi^+$ & 17.7& 9.2 $\pm$ 0.6\\
\hline
 $\Dstarp\to\Dz\pip$;& &\\
\ \ \ \  $\Dz \rightarrow K^{-}\pi^{+}$& 18.5                       &2.57 $\pm$ 0.06\\
\ \ \ \  $\Dz \rightarrow K^{-}\pi^{+}\piz$ & 6.4                   & 8.8 $\pm$ 0.5\\
\ \ \ \  $\Dz \rightarrow K^{-}\pi^{+}\pi^{-}\pi^{+}$&10.1          &5.05 $\pm$ 0.21\\
\ \ \ \  $\Dz \rightarrow \KS\pi^{+}\pi^{-}$; $\KS\to\pim\pip$ & 10.6  &1.37 $\pm$ 0.08\\
\hline
\end{tabular}}
\end{center}
\label{tab:eff}
\end{table}

After the  selection described above, two classes of backgrounds remain. First, there is {\it{combinatorial background}} in the signal region, coming from random combinations of 
tracks in the event. 
We estimate this background from the sideband of 
the $m_{ES}$ distribution, describing it with a threshold function 
$dN/dm_{ES}\propto m_{ES}\sqrt{1-m_{ES}^{2}/E_b^{*2}}\exp\left[-\xi\left(1-m_{ES}^{2}/E_b^{*2}\right)\right]$, characterized by the shape parameter $\xi$~\cite{argus}.  We obtain the parameter $\xi$  from a
 fit to the distributions of $m_{ES}$ in data, in the $\Delta E$ sideband region. The number of combinatorial background events is obtained by scaling the events in the sideband of the $m_{ES}$ distribution into the signal region with the ratio of the threshold function area in the two regions. Including systematic errors, we estimate 56.3 $\pm$ 3.0 and 22.0 $\pm$ 1.8 events for the ${\ensuremath {\Bp{\to}D^{+}K^0_s}}$  and ${\ensuremath {\Bp{\to}D^{*+}K^0_s}}$ mode, respectively.
Second, there is $peaking$ $background$ due to misreconstructed $B$ meson 
decays that have an $m_{ES}$ distribution peaking near the $B$ mass. We 
study the peaking background with MC and we estimate it to be 4.4 $\pm$ 
1.2  and 1.2 $\pm$ 0.6 events for the ${\ensuremath {\Bp{\to}D^{+}K^0_s}}$  
and ${\ensuremath {\Bp{\to}D^{*+}K^0_s}}$ modes, respectively. The
dominant contribution to the peaking background comes from well-known 
$B^0 \rightarrow D^{(*)-}\X^+$ decays ($X^+$ = $\pi^+$, $\rho^+$, $a^+_1$).
As a cross-check, we also estimate the peaking background using
candidates from the $D$ mass sidebands in data and we find results
consistent with the MC prediction.

\begin{figure}
\begin{center}
\includegraphics[width=8cm]{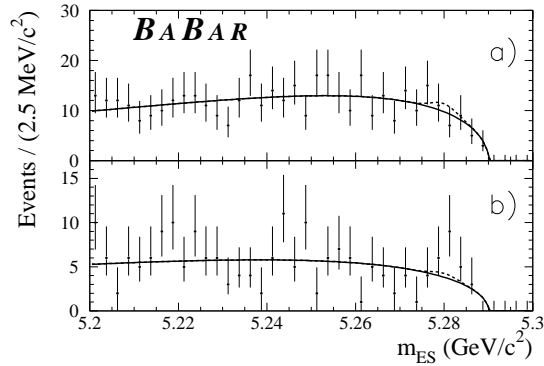}
\caption{The \mes\ distribution for the a) ${\ensuremath {\Bp{\to}D^{+}K^0_s}}$  and b)  ${\ensuremath {\Bp{\to}D^{*+}K^0_s}}$
 candidates within the $\Delta E$ signal band in data  after all selection 
requirements. Combinatorial (full line) and peaking  (dashed line)  backgrounds are superimposed.
\label{fig:mesb}
}
\end{center}
\end{figure}

Figure~\ref{fig:mesb} shows the $m_{ES}$ distributions in the $\Delta E$ signal band
 for the two modes after the selection. The expected background is superimposed.
\begin{table}[b]
\caption{The number of candidates  in the signal region in data ($N_{\rm{cand}}$), the corresponding expected combinatorial background ($N_{\rm{comb}}$), the peaking background  ($N_{\rm{peak}}$),
the probability ($P_{bkgd}$) of the data being consistent with the background fluctuating up to the level of the data in absence of signal, and the 90$\%$ confidence-level upper limit. Systematic uncertainties are included.}
\begin{center}
\mbox{ \small
\begin{tabular}{l|c|c|c||c|c} \hline \hline
$B$ mode& $N_{\rm{cand}}$ & $N_{\rm{comb}}$  & $N_{\rm{peak}}$ & $P_{bkgd}(\%)$ & 90$\%$ C.L. \\
\hline
${D^+}\ K^0_{s}$     & 57  & 56.3 $\pm$ 3.0 &  4.4 $\pm$ 1.2& 69 & $\dclim$ \\
\hline
${D^{*+}}\ K^0_{s}$ & 28& 22.0 $\pm$ 1.8 &  1.2 $\pm$ 0.6 & 24 & $\dstarclim$ \\
\hline
\end{tabular}}
\end{center}
\label{tab:fit}
\end{table}

To compute the confidence level (C.L.)  at which the data agree with
a given hypothesis on ${\ensuremath {{\cal B}(\Bp{\to}D^{(*)+}K^0})}$  we 
 use a frequentist technique~\cite{freq}, which
  treats properly the small number of events and includes the systematic errors directly 
in the computation of confidence
 intervals or limits.
The C.L. is defined as the fraction of times a random number, following the expected 
distribution of the number of events 
in the signal region ($N_{exp}$), exceeds the number of observed events ($N_{cand}$ 
in Tab.~\ref{tab:fit}).
 $N_{exp}$ is distributed according to the sum of Poissonian distributions  with mean 
values $\mu$ distributed as follows: 
for a given value of   ${\ensuremath {{\cal B}(\Bp{\to}D^{(*)+}K^0})}$  we estimate
 $\mu$ as the sum of the expectation value of the number 
of events from  the combinatorial and peaking  background ( $N_{comb}$ and 
$N_{peak}$, respectively), and from the signal ($N_{sig}$), 
$\mu = N_{comb}+N_{peak}+N_{sig}$.

We estimate $N_{comb}$ by scaling the number of events in the $m_{ES}$ sideband to 
the signal region and
by considering the Poisson fluctuations of the number of events in the sideband and 
the systematic uncertainties on the threshold parameter $\xi$.
 We estimate $N_{peak}$ from the MC, taking into account its limited statistics. 
Table~\ref{tab:fit} reports the mean values and standard deviations for $N_{comb}$ and $N_{peak}$.
Finally, for a given value of the branching fraction, $N_{sig}$ is obtained as:
\begin{equation}
N_{sig} = {\cal B}(B^+ \rightarrow D^{(*)+}K^0) \times N_{B} \times \Sigma_i \epsilon_i {\cal B}_i
\end{equation}
where  the number of $B^{\pm}$ mesons ($N_{B}$) and the product of the efficiency and the branching  fraction of the 
sub-decay modes  ($\Sigma_i \epsilon_i {\cal B}_i$) are varied according to Gaussian distributions 
within their systematic uncertainties.
The systematic errors on the reconstruction efficiency are shown in Table~\ref{tab:sys} and include the 
uncertainty due to limited MC statistics, uncertainty on tracking 
efficiency, $K^0_s$ and $\pi^0$ reconstruction, charged-kaon identification, other selection criteria.
They have all been estimated by comparing the data and simulation performances in control samples. 
Also, the uncertainties on $N_B$ ($1.1\%$) and  on the branching fraction of the sub-decay modes have been taken into account.
The total uncertainty is obtained by adding the  contributions from the individual sources in quadrature.
\begin{table*}[b]
\caption{Relative systematic errors on the branching fraction due to, 
respectively: 
MC statistics , track reconstruction , Kaon identification,
\KS  and \piz  reconstruction efficiencies, and  the data-MC agreement on the signal shapes of  
$\Delta E$, cos$\theta_{T}$, and $\cal{F}$. }
\begin{center}
\mbox{ \small
\begin{tabular}{l|c|c|c|c|c|c|c|c|c} \hline \hline
$D$ mode& MC($\%$) & Tracks($\%$)  & Kaon($\%$) & $K^0_s$($\%$) & $\pi^0$$(\%)$ &  $\Delta E$$(\%)$ & cos$\theta_{T}$$(\%)$ &$\cal{F}$($\%$) & Total($\%$)\\
& & & & & &  & & &  \\ \hline
\ \ $D^+ \rightarrow K^0_s\pi^+$ & 1.4 & 1.1 &- & 0.3& - &0.4 & 0.4 & 0.7&2.0\\
\ \ $D^+ \rightarrow K^-\pi^+\pi^+$ & 0.5& 1.1& 0.4& 0.2 & - & 0.4& 0.4& 0.7& 1.5\\
\hline
\ \ $\Dz \rightarrow K^{-}\pi^{+}$& 0.9& 1.2& 0.4& 0.2& - & 0.8&0.4& 0.7& 1.8\\
\ \ $\Dz \rightarrow K^{-}\pi^{+}\piz$ & 0.3& 0.4&0.1 & 0.1& 0.3&0.2&0.1& 0.3&0.7\\
\ \ $\Dz \rightarrow K^{-}\pi^{+}\pi^{-}\pi^{+}$ & 0.5&0.9 &0.2 & 0.1 &  -& 0.1&0.2& 0.4& 1.2\\
\ \ $\Dz \rightarrow \KS\pi^{+}\pi^{-}$ & 1.0& 1.0& -& 0.2& - & 0&0.2&0.4 &1.5\\
\hline
\end{tabular}}
\end{center}
\label{tab:sys}
\end{table*}

Calculating the C.L. with the procedure just described and setting ${\ensuremath {{\cal B}(\Bp{\to}D^{(*)+}K^0})}=0$,  we estimate
the probability of the background to fluctuate above  the observed number of events to be
 69$\%$ and 24$\%$ for the ${\ensuremath {\Bp{\to}D^{+}K^0_s}}$ and for the ${\ensuremath {\Bp{\to}D^{*+}K^0_s}}$ modes, respectively.
In absence of significant signal
we then set the following upper limits on the values of the branching fractions corresponding to a C.L. of $ 90\%$:\\
\begin{eqnarray}
{\ensuremath {{\cal B}(\Bp{\to}D^{+}K^0})}&<& \dclim , \\
{\ensuremath {{\cal B}(\Bp{\to}D^{*+}K^0})} &<&\dstarclim .
\nonumber
\end{eqnarray}
 We also compute the branching fractions
${\cal B}(\Bp{\to}D^{+}K^0) = (-0.28^{+0.61}_{-0.56})\times 10^{-5}$ and 
$ {\cal B}(\Bp{\to}D^{*+}K^0) = (0.28^{+0.44}_{-0.41})\times 10^{-5}$.
The errors above include both the statistical and the systematic uncertainties.

In conclusion, we report on the search for the rare decays ${\ensuremath {\Bp{\to}D^{(*)+}K^0_s}}$, 
which are predicted to proceed through annihilation diagrams.
We do not observe any significant signal and we set  90\% C.L. upper limits on their branching fractions.

We are grateful for the excellent luminosity and machine conditions
provided by our \pep2\ colleagues, 
and for the substantial dedicated effort from
the computing organizations that support \babar.
The collaborating institutions wish to thank 
SLAC for its support and kind hospitality. 
This work is supported by
DOE
and NSF (USA),
NSERC (Canada),
IHEP (China),
CEA and
CNRS-IN2P3
(France),
BMBF and DFG
(Germany),
INFN (Italy),
FOM (The Netherlands),
NFR (Norway),
MIST (Russia), and
PPARC (United Kingdom). 
Individuals have received support from CONACyT (Mexico), A.~P.~Sloan Foundation, 
Research Corporation,
and Alexander von Humboldt Foundation.


\begin{thebibliography}{99}
\bibitem{ch}
Charge conjugation is implied throughout this letter.
\end{thebibliography}
\end{document}